\documentclass[twocolumn,trackchanges]{aastex701}

\usepackage{booktabs}
\usepackage{orcidlink}
\usepackage{multirow}
\usepackage{makecell}
\usepackage{seqsplit}
\usepackage{color, xcolor}

\begin{document}

\title{Filtering Interlopers with Photometry and Diagnostic Features: A Machine Learning Framework Validated with CSST Slitless Spectroscopy}

\author[orcid=0000-0003-3929-3229, gname=Hui, sname=Peng]{Hui Peng}
\affiliation{State Key Laboratory of Dark Matter Physics, Shanghai Jiao Tong University, Shanghai 200240, People’s Republic of China}
\affiliation{Department of Astronomy, School of Physics and Astronomy, Shanghai Jiao Tong University, Shanghai 200240, People’s Republic of China}
\affiliation{Key Laboratory for Particle Astrophysics and Cosmology (MOE)/Shanghai Key Laboratory for Particle Physics and Cosmology, Shanghai Jiao Tong University, Shanghai 200240, People’s Republic of China}
\email[show]{huipeng@sjtu.edu.cn}

\author[orcid=0000-0002-9359-7170, gname=Yu, sname=Yu]{Yu Yu}
\affiliation{State Key Laboratory of Dark Matter Physics, Shanghai Jiao Tong University, Shanghai 200240, People’s Republic of China}
\affiliation{Department of Astronomy, School of Physics and Astronomy, Shanghai Jiao Tong University, Shanghai 200240, People’s Republic of China}
\affiliation{Key Laboratory for Particle Astrophysics and Cosmology (MOE)/Shanghai Key Laboratory for Particle Physics and Cosmology, Shanghai Jiao Tong University, Shanghai 200240, People’s Republic of China}
\email[show]{yuyu22@sjtu.edu.cn}

\author[gname=Yiyang, sname=Guo]{Yiyang Guo}
\affiliation{State Key Laboratory of Dark Matter Physics, Shanghai Jiao Tong University, Shanghai 200240, People’s Republic of China}
\affiliation{Department of Astronomy, School of Physics and Astronomy, Shanghai Jiao Tong University, Shanghai 200240, People’s Republic of China}
\affiliation{Key Laboratory for Particle Astrophysics and Cosmology (MOE)/Shanghai Key Laboratory for Particle Physics and Cosmology, Shanghai Jiao Tong University, Shanghai 200240, People’s Republic of China}
\email{yiyang_guo@outlook.com}

\author[orcid=0000-0003-3196-7938, gname=Yizhou, sname=Gu]{Yizhou Gu}
\affiliation{State Key Laboratory of Dark Matter Physics, Shanghai Jiao Tong University, Shanghai 200240, People’s Republic of China}
\affiliation{Tsung-Dao Lee Institute, and Key Laboratory for Particle Physics, Astrophysics and Cosmology, Ministry of Education, Shanghai Jiao Tong University, Shanghai 201210, People’s Republic of China}
\email{guyizhou@sjtu.edu.cn}

\author[orcid=0000-0002-8705-6327, gname=Run, sname=Wen]{Run Wen}
\affiliation{State Key Laboratory of Dark Matter Physics, Shanghai Jiao Tong University, Shanghai 200240, People’s Republic of China}
\affiliation{Tsung-Dao Lee Institute, and Key Laboratory for Particle Physics, Astrophysics and Cosmology, Ministry of Education, Shanghai Jiao Tong University, Shanghai 201210, People’s Republic of China}
\email{wenrun1214@sjtu.edu.cn}

\author[orcid=0000-0002-2547-0434, gname=Yunkun, sname=Han]{Yunkun Han}
\affiliation{Yunnan Observatories, Chinese Academy of Sciences, 396 Yangfangwang, Guandu District, Kunming 650216, People’s Republic of China}
\email{hanyk@ynao.ac.cn}

\author[orcid=0000-0003-2132-0727, gname=Jipeng, sname=Sui]{Jipeng Sui}
\affiliation{Key Laboratory of Optical Astronomy, National Astronomical Observatories, Chinese Academy of Sciences, Beĳing 100101, People’s Republic of China}
\affiliation{School of Astronomy and Space Science, University of Chinese Academy of Sciences, Beĳing 101408, People’s Republic of China}
\email{suijp@bao.ac.cn}

\author[orcid=0000-0002-6684-3997, gname=Hu, sname=Zou]{Hu Zou}
\affiliation{Key Laboratory of Optical Astronomy, National Astronomical Observatories, Chinese Academy of Sciences, Beĳing 100101, People’s Republic of China}
\affiliation{School of Astronomy and Space Science, University of Chinese Academy of Sciences, Beĳing 101408, People’s Republic of China}
\email{zouhu@nao.cas.cn}

\author[orcid=0000-0003-3997-4606, gname=Xiaohu, sname=Yang]{Xiaohu Yang}
\affiliation{State Key Laboratory of Dark Matter Physics, Shanghai Jiao Tong University, Shanghai 200240, People’s Republic of China}
\affiliation{Tsung-Dao Lee Institute, and Key Laboratory for Particle Physics, Astrophysics and Cosmology, Ministry of Education, Shanghai Jiao Tong University, Shanghai 201210, People’s Republic of China}
\affiliation{Department of Astronomy, School of Physics and Astronomy, Shanghai Jiao Tong University, Shanghai 200240, People’s Republic of China}
\affiliation{Key Laboratory for Particle Astrophysics and Cosmology (MOE)/Shanghai Key Laboratory for Particle Physics and Cosmology, Shanghai Jiao Tong University, Shanghai 200240, People’s Republic of China}
\email{xyang@sjtu.edu.cn}

\author[orcid=0000-0003-2632-9915, gname=Pengjie, sname=Zhang]{Pengjie Zhang}
\affiliation{State Key Laboratory of Dark Matter Physics, Shanghai Jiao Tong University, Shanghai 200240, People’s Republic of China}
\affiliation{Department of Astronomy, School of Physics and Astronomy, Shanghai Jiao Tong University, Shanghai 200240, People’s Republic of China}
\affiliation{Tsung-Dao Lee Institute, and Key Laboratory for Particle Physics, Astrophysics and Cosmology, Ministry of Education, Shanghai Jiao Tong University, Shanghai 201210, People’s Republic of China}
\affiliation{Key Laboratory for Particle Astrophysics and Cosmology (MOE)/Shanghai Key Laboratory for Particle Physics and Cosmology, Shanghai Jiao Tong University, Shanghai 200240, People’s Republic of China}
\email{zhangpj@sjtu.edu.cn}

\author[orcid=0000-0003-3728-9912, gname=Xian Zhong, sname=Zheng]{Xian Zhong Zheng}
\affiliation{State Key Laboratory of Dark Matter Physics, Shanghai Jiao Tong University, Shanghai 200240, People’s Republic of China}
\affiliation{Tsung-Dao Lee Institute, and Key Laboratory for Particle Physics, Astrophysics and Cosmology, Ministry of Education, Shanghai Jiao Tong University, Shanghai 201210, People’s Republic of China}
\email{xzzheng@sjtu.edu.cn}

\author[orcid=0000-0003-4936-8247, gname=Hong, sname=Guo]{Hong Guo}
\affiliation{Key Laboratory for Research in Galaxies and Cosmology,
Shanghai Astronomical Observatory, Shanghai 200030, People’s Republic of China}
\email{guohong@shao.ac.cn}

\author[orcid=0000-0002-4534-3125, gname=Yipeng, sname=Jing]{Yipeng Jing}
\affiliation{State Key Laboratory of Dark Matter Physics, Shanghai Jiao Tong University, Shanghai 200240, People’s Republic of China}
\affiliation{Tsung-Dao Lee Institute, and Key Laboratory for Particle Physics, Astrophysics and Cosmology, Ministry of Education, Shanghai Jiao Tong University, Shanghai 201210, People’s Republic of China}
\affiliation{Department of Astronomy, School of Physics and Astronomy, Shanghai Jiao Tong University, Shanghai 200240, People’s Republic of China}
\affiliation{Key Laboratory for Particle Astrophysics and Cosmology (MOE)/Shanghai Key Laboratory for Particle Physics and Cosmology, Shanghai Jiao Tong University, Shanghai 200240, People’s Republic of China}
\email{ypjing@sjtu.edu.cn}

\author[orcid=0000-0002-8711-8970, gname=Cheng, sname=Li]{Cheng Li}
\affiliation{Department of Astronomy, Tsinghua University, Beĳing 100084, People’s Republic of China}
\email{cli2015@mail.tsinghua.edu.cn}

\author[gname=Hu, sname=Zhan]{Hu Zhan}
\affiliation{National Astronomical Observatories, Chinese Academy of Sciences, 20A Datun Road, Beijing 100101, People’s Republic of China}
\affiliation{Kavli Institute for Astronomy and Astrophysics, Peking University, Beijing 100871, People’s Republic of China}
\email{zhanhu@nao.cas.cn}

\author[orcid=0000-0003-4726-6714, gname=Gongbo, sname=Zhao]{Gongbo Zhao}
\affiliation{National Astronomical Observatories, Chinese Academy of Sciences, 20A Datun Road, Beijing 100101, People’s Republic of China}
\affiliation{School of Astronomy and Space Sciences, University of Chinese Academy of Sciences (UCAS), 19A Yuquan Road, Beijing 100049, People’s Republic of China}
\email{gbzhao@nao.cas.cn}

\correspondingauthor{Hui Peng, Yu Yu}

\begin{abstract}
The slitless spectroscopic method employed by missions such as Euclid and the Chinese Space-station Survey Telescope (CSST) faces a fundamental challenge: spectroscopic redshifts derived from their data are susceptible to emission-line misidentification due to the limited spectral resolution and signal-to-noise ratio.
This effect systematically introduces interloper galaxies into the sample.
Conventional strict selection not only struggles to secure high redshift purity but also drastically reduces completeness by discarding valuable data.
To overcome this limitation, we develop an XGBoost classifier that leverages photometric properties and spectroscopic diagnostics to construct a high-purity redshift catalog while maximizing completeness.
We validate this method on a simulated sample with spectra generated by the CSST emulator for slitless spectroscopy.
Of the $\sim$62 million galaxies that obtain valid redshifts (parent sample), approximately 43\% achieve accurate measurements, defined as $|\Delta z| \leqslant 0.002(1+z)$.
From this parent sample, the XGBoost classifier selects galaxies with a selection efficiency of 42.3\% on the test set and 42.2\% when deployed on the entire parent sample.
Crucially, among the retained galaxies, 96.6\% (parent sample: 96.5\%) achieve accurate measurements, while the outlier fraction ($|\Delta z|>0.01(1+z)$) is constrained to 0.13\% (0.11\%).
We verified that simplified configurations that exclude either spectroscopic diagnostics (except the measured redshift) or photometric data yield significantly higher outlier fractions, increasing by factors of approximately 3.5 and 6.3, respectively, with the latter case also introducing notable catastrophic interloper contamination.
This framework effectively resolves the purity-completeness trade-off, enabling robust large-scale cosmological studies with CSST and similar surveys.

\end{abstract}


\keywords{\uat{Redshift surveys}{1378} --- \uat{Astronomy data analysis}{1858} --- \uat{Large-scale structure of the universe}{902}}


\section{Introduction}\label{sec:intro}
Spectroscopic surveys are foundational to modern cosmology, enabling the systematic study of galaxy formation and evolution, the precise mapping of the large-scale structure (LSS) of our Universe, and stringent constraints on dark matter and dark energy.
To map these structures over increasingly vast volumes, a new generation of ongoing and upcoming surveys is now expanding our observational horizons.
Notable projects include the Dark Energy Spectroscopic Instrument \citep[DESI;][]{DESI-Collaboration:2025aa},  Euclid \citep{Euclid-Collaboration:2025ai}, SPHEREx \citep{Bock:2025aa}, the Hobby–Eberly Telescope Dark Energy Experiment \citep[HETDEX;][]{Gebhardt:2021aa}, the Nancy Grace Roman Space Telescope \citep[Roman;][]{Spergel:2015aa}, and the Chinese Space-station Survey Telescope \citep[CSST;][]{CSST-Collaboration:2026aa}.

The scientific return of these surveys hinges on the accurate measurement of spectroscopic redshifts.
This measurement is determined by comparing the observed wavelength ($\lambda_{\text{obs}}$) of a spectral feature with its intrinsic rest-frame wavelength ($\lambda_{\text{rest}}$), according to the relation $1+z = \lambda_{\text{obs}} / \lambda_{\text{rest}}$.
In high-resolution or high signal-to-noise ratio (SNR) observations, the concurrent detection of multiple emission lines with fixed and known wavelength separations enables a secure and unambiguous redshift measurement.

However, low-resolution slitless spectroscopic surveys, including those conducted by Roman, Euclid, and CSST, introduce significant challenges.
These arise primarily from limited spectral resolution, low typical SNR of detected emission lines, and contamination from adjacent objects \citep{Wen:2024ab,Euclid-Collaboration:2025ab,Bock:2025aa}.
Under these conditions, emission lines become susceptible to misidentification, where a detected feature may be incorrectly matched to a rest-frame wavelength, leading to interlopers.
Such a feature could correspond not to the targeted emission line, but to a different line or even a prominent noise spike \citep{Euclid-Collaboration:2025ac,Euclid-Collaboration:2025ak}.
These misidentifications systematically introduce interlopers into the galaxy sample, which in turn biases the measured clustering statistics and ultimately compromises the accuracy of inferred cosmological parameters \citep{Pullen:2016vt}.

Research has shown that many fundamental cosmological statistics and probes, such as the power spectrum, correlation function, redshift-space distortions, baryon acoustic oscillations, and weak lensing, can be significantly affected by the presence of interlopers \citep{Pullen:2016vt,Leung:2017aa,Addison:2019aa,Grasshorn-Gebhardt:2019aa,Awan:2020aa,Massara:2021aa,Hilmi:2024aa,Euclid-Collaboration:2025ad}.
While mitigating this contamination is crucial, conventional approaches often impose strict selection criteria that discard a significant fraction of observations \citep{Euclid-Collaboration:2025ac,Sui:2025aa}.
This approach not only struggles to guarantee a high-purity redshift catalog but also drastically reduces sample completeness, thereby discarding valuable data and weakening statistical power.
To overcome this inherent limitation, recent methodologies have diverged into two complementary paths.
The first strategy focuses on constructing high-purity catalogs by incorporating photometric data sources.
This can be realized in two distinct ways: by statistically defining a pure sample from slitless spectroscopic observations using photometric cuts \citep{Cagliari:2024aa}, or by resolving ambiguous cases (e.g., single-line galaxies) through a combination of photometric data (or derived properties like photometric equivalent widths) with early neural networks \citep{Kirby:2007aa} or Bayesian frameworks \citep{Leung:2017aa,Davis:2023aa}.
The second strategy aims to directly quantify the interloper fraction and recover the true target signal even when contamination exists.
This is commonly achieved through the analysis of measured correlation functions and power spectra \citep{Grasshorn-Gebhardt:2019aa,Farrow:2021aa,Gong:2021tb,Foroozan:2022aa, Peng:2023aa, Nguyen:2024aa, Bernal:2025aa} or by employing machine learning methods such as neural networks \citep{Massara:2023aa,Cagliari:2025aa}.

\begin{figure*}[htbp]
    \centering
	\includegraphics[width=16cm]{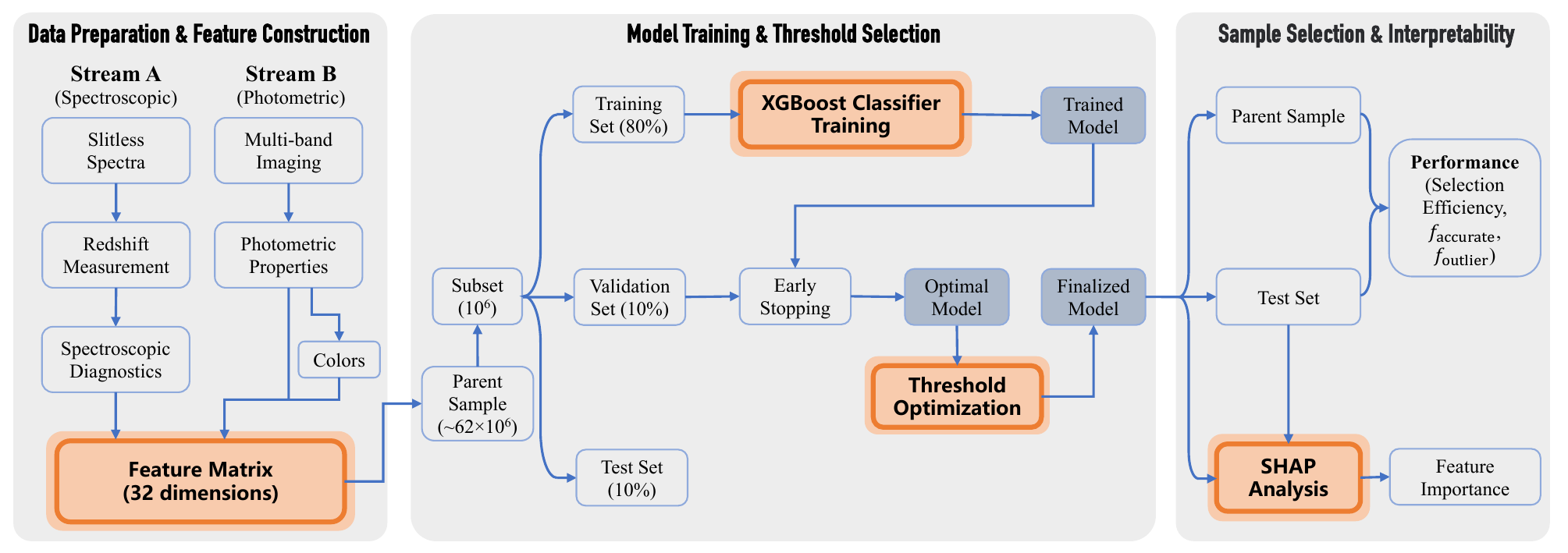}
    \caption{Schematic workflow of the machine learning framework for constructing a high-purity, high-completeness galaxy sample. The diagram illustrates the pipeline from multi-source data preparation and feature engineering (combining photometric properties with spectroscopic diagnostics), through XGBoost classifier training and threshold optimization, to final evaluation and SHAP-based interpretability analysis. The output catalog is optimized for both high selection efficiency and high accuracy (high $f_{\mathrm{accurate}}$ and low $f_{\mathrm{outlier}}$ defined in Section~\ref{sec:model_performance}).}
    \label{fig:schematic_overview}
\end{figure*}

Building upon the first strategy as a whole, we present a novel XGBoost classification framework that leverages multiband photometric data and key spectroscopic diagnostics to simultaneously achieve high purity and high completeness in low-resolution slitless spectroscopy.
We systematically validate this model using mock data generated specifically for CSST and employ SHapley Additive exPlanations (SHAP) analysis to interpret the model and quantify feature importance.
The overall workflow of the proposed framework is illustrated in Figure~\ref{fig:schematic_overview}, encompassing the integration of multiband photometry and spectroscopic diagnostics, the training and optimization of the XGBoost classifier, and the final evaluation of the selected sample.
While validated in the CSST context, the framework is inherently general and can be adapted to other low-resolution spectroscopic surveys facing similar challenges.

The remainder of this paper is structured as follows.
Section~\ref{sec:simulated_dataset} details the CSST survey and the mock dataset employed for validation.
Section~\ref{sec:method} describes our full selection methodology, covering both the XGBoost model training and the subsequent threshold selection strategy.
Section~\ref{sec:results} presents the performance of our fiducial model, along with a SHAP-based interpretability analysis, and compares results obtained with simplified feature sets.
Section~\ref{sec:conclusions} summarizes our findings and discusses their implications for future work.

\section{The Simulated Dataset}\label{sec:simulated_dataset}
\subsection{CSST Slitless Spectroscopic Survey Simulation}\label{sec:CSST_simulation}
CSST is scheduled to be launched in 2027, which will conduct simultaneous multiband photometric and slitless spectroscopic surveys \citep{Zhan:2021aa}.
The main focal plane of the survey camera is divided into seven photometric imaging bands ($NUV$, $u$, $g$, $r$, $i$, $z$, and $y$) and three slitless spectroscopic bands ($GU$, $GV$, and $GI$; average spectral resolution $R=\lambda/\Delta\lambda\geqslant$ 200), covering a wavelength range of 255–1000 nm \citep{Cao:2018aa}.
The survey plans to perform approximately 7 yr of cumulative observations within its 10 yr orbital period, acquiring both wide-field data over \(17{,}500\ \mathrm{deg}^2\) and deep-field data over \(400\ \mathrm{deg}^2\) in both photometric and slitless spectroscopic bands \citep{CSST-Collaboration:2026aa}.
With hundreds of millions of spectra and over one billion photometric data of galaxies collected, it is expected to play an important role in both cosmological studies and a wide range of astronomical research fields \citep{Gong:2025aa}.

As a benchmark for investigating various selection effects in future CSST observations, the reference mock galaxy redshift survey (MGRS) that an ideal spectroscopic survey would obtain has been constructed in \citet{Gu:2024aa}, with a redshift range of $0 < z < 1$ and a magnitude limit of $z$-band $< 21$.
The mock catalog is constructed  based on one high-resolution $N$-body simulation from the Jiutian simulation suite \citep{Han:2025aa}, which is designed to support the scientific analysis of the CSST extragalactic surveys.
The assignment of galaxy luminosities to the central and subhalo populations is performed via the subhalo abundance matching (SHAM) method, utilizing the DESI $z$-band luminosity functions at different redshifts \citep{Dey:2019aa}.
Each galaxy in the sample is matched with its counterpart in the DESI Legacy Imaging Surveys Data Release 9 (LS DR9) group catalog based on redshift, luminosity, and halo mass \citep{Yang:2021vo}.

To quantitatively assess the survey's potential, the CSST Emulator for Slitless Spectroscopy (CESS) has been developed to efficiently generate simulated 1D slitless spectra \citep{Wen:2024ab}.
By applying CESS to the simulated galaxies in the MGRS, a galaxy spectrum library containing approximately $1.4 \times 10^8$ galaxies has been generated, which includes best-fit spectra, redshifts, sky coordinates, and physical parameters.
The simulated magnitudes in the bands are generated by convolving galaxy spectral energy distributions (SEDs) with the filter transmission curves and sensitivity specifications of the CSST survey camera.
The analysis of the simulated CSST spectra indicates that CSST slitless spectroscopy can potentially provide reliable redshifts \citep[directly obtained from the spectra using three criteria in][]{Wen:2024ab} for approximately one-quarter of the galaxies in the sample, while the remainder either lack emission-line features or are not bright enough for detection.

Subsequently, \citet{Sui:2025aa} developed a redshift measurement method for slitless spectroscopy in the upcoming CSST survey, which operates by identifying emission lines in the observed spectra and matching them to their corresponding rest-frame wavelengths.
Although an alternative deep learning-based method was proposed by \citet{Zhou:2024aa}, we employ the approach from \citet{Sui:2025aa} in this analysis.
It utilizes six emission lines for redshift determination: $\mathrm{C\,\scriptstyle{III}}]\lambda1909$, $[\mathrm{O\,\scriptstyle{II}}]\lambda3727$, $\mathrm{H\gamma}$, $\mathrm{H\beta}$, $[\mathrm{O\,\scriptstyle{III}}]\lambda\lambda4959, 5007$, and $\mathrm{H\alpha}$.
The redshift measurement process can output a suite of diagnostic information, including quality flags (ZWARNING), the number of matched lines, and line properties.
However, applying thresholds directly to these diagnostics (e.g., requiring a minimum number of matched lines or specific ZWARNING values) creates a fundamental trade-off: this strict, rule-based filtering inevitably discards a substantial fraction of observations, thereby preventing the construction of a redshift catalog that is both statistically large (high completeness) and highly pure.

\subsection{Mock Catalog}\label{sec:mock_catalog}
\begin{figure*}[htbp]
    \centering
	\includegraphics[width=16.5cm]{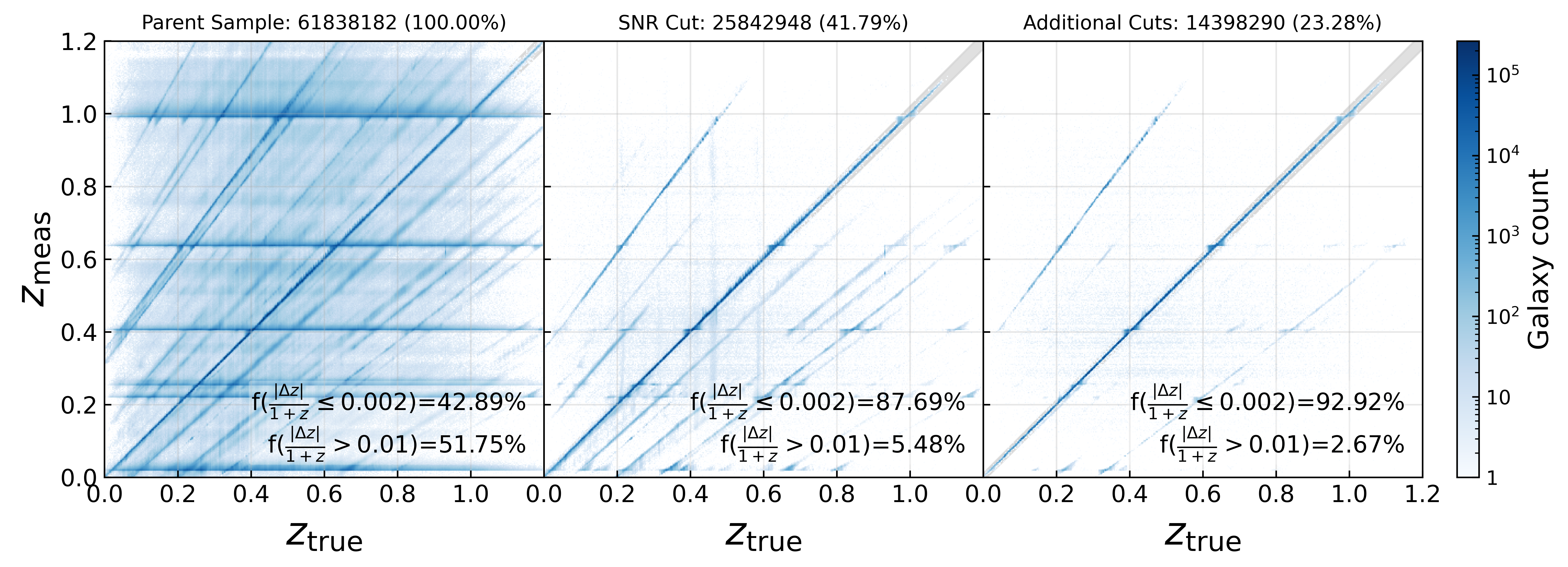}
    \caption{Comparison of measured versus true redshifts for galaxies with valid redshift measurements under different selection criteria. Left: parent sample of galaxies with valid redshifts. Middle: traditional SNR‑based selection: $\mathrm{H\alpha}_{\rm snr}\geqslant5\,{\rm OR}\, ([\mathrm{O\,\scriptstyle{III}}]_{\rm snr}\geqslant5\,{\rm AND}\,\mathrm{H\beta}_{\rm snr}\geqslant 5)$.  Right: traditional combined‑criteria selection: the above SNR criteria plus ZWARNING = 0 and $ N_{\rm lines}\geqslant3$. In each panel, the gray shaded regions denote the boundaries beyond which objects are classified as outliers ($|\Delta z| > 0.01(1+z_{\rm{true}})$). The title in each panel shows the total galaxy count and the selection efficiency (i.e., the fraction of selected objects relative to the total sample). The corresponding fractions of accurate measurements ($|\Delta z| \leqslant 0.002(1+z_{\rm{true}})$) and outliers are explicitly annotated.
    }
    \label{fig:sample_cut_compare}
\end{figure*}

We applied the redshift measurement algorithm to the complete set of simulated 1D slitless spectra in \citet{Wen:2024ab}.
This yielded a total of 137,855,710 galaxies with the requisite output data, of which 61,838,182 ($\sim$45\%) provided valid redshift measurements ($z_{\mathrm{meas}} > 0$).
These galaxies constitute the parent sample for all subsequent analysis. 
Within this parent sample, 42.9\% achieve accurate measurements, satisfying the criterion $|\Delta z| \leqslant 0.002(1+z_{\mathrm{true}})$, where $\Delta z = z_{\mathrm{meas}} - z_{\mathrm{true}}$ \citep{CSST-Collaboration:2026aa}.

Figure~\ref{fig:sample_cut_compare} compares the measured‑versus‑true redshift distributions for three samples: the parent catalog with all valid redshifts (left), a subset selected by traditional SNR criteria (middle), and a further‑refined subset using combined traditional cuts (right).
The traditional SNR cuts require $\mathrm{H\alpha}_{\rm snr}\geqslant5\,{\rm OR}\, ([\mathrm{O\,\scriptstyle{III}}]_{\rm snr}\geqslant5\,{\rm AND}\,\mathrm{H\beta}_{\rm snr}\geqslant 5)$.
The combined cuts add the conditions ZWARNING = 0 and $ N_{\rm lines}\geqslant3$ to the SNR criteria above.
According to the detailed analysis of emission-line redshift measurement in \citet{Sui:2025aa}, the interloper population in the parent sample originates from three distinct mechanisms: (1) Instrumental artifacts: The clustering of redshifts at specific values is primarily caused by the misidentification of grating boundaries where emission lines fall into the low-SNR overlap regions between adjacent spectral bands \citep[see Figure 1 in][]{Sui:2025aa}.
(2) Emission-line confusion: The off-diagonal linear aggregations arise from systematic degeneracies in line matching, including misidentified genuine line pairs (e.g., $\mathrm{H\alpha}$ + $[\mathrm{O\,\scriptstyle{III}}]$ mistaken for $\mathrm{H\beta}$ + $[\mathrm{O\,\scriptstyle{II}}]$) and mismatched combinations of a real line with one or more spurious features---for instance, $\mathrm{H\alpha}$ being mistaken for $\mathrm{H\gamma}$ alongside a false detection.
(3) Spurious detections: The chaotic scatter of points stems from noise spikes in low-SNR spectra, or from continuum misestimation induced by absorption features in bright quiescent galaxies, which generates spurious emission-line candidates.
Even with the combined cuts, which involve a significant loss of valuable galaxies (selection efficiency of only 23.28\%), the fraction of accurate measurements remains unsatisfactory at 92.92\%.
The comparison across these panels reveals that conventional selection criteria are insufficient to achieve both high redshift purity and completeness.

To enable computationally efficient model development and to match the typical scale of high‑resolution spectroscopic reference samples (e.g., from DESI), we constructed a representative random subsample of $10^6$ sources from the parent catalog ($\sim$62 million galaxies).
This subsample was partitioned into training (80\%), validation (10\%), and test (10\%) sets for the subsequent analysis.

To systematically identify interlopers, we employ 32 features derived from two categories.
First, from the slitless spectra themselves, we extract 10 spectroscopic diagnostics originating from the redshift measurement process.
Second, from the imaging data, we use apparent magnitudes in 10 photometric bands (including the three slitless bands $GU$, $GV$, and $GI$) and 4 morphological parameters (Sérsic index, effective radius, position angle, and axis ratio).
The 24 base features are listed in Table~\ref{tab:base_quantities}.
In addition, we compute eight colors from the photometric bands ($NUV-u$, $u-g$, $g-r$, $r-i$, $i-z$, $z-y$, $GU-GV$, and $GV-GI$).
This combined feature set enables the clear identification of interlopers.

While our current model is trained on CSST‑specific features, we note that other slitless spectroscopy surveys (e.g., Euclid, Roman) provide analogous photometric properties and spectroscopic diagnostics.
Although filter curves and spatial resolutions differ, the physical information content (emission line strengths, multi‑band fluxes, and morphological parameters) remains similar.
Therefore, we expect the model to be transferable with only minor degradation.
A quantitative assessment of cross‑survey performance is left for future work.

\begin{table*}[ht]
    \centering
    \caption{The 24 base features used for interloper identification.}
    \label{tab:base_quantities}
    \begin{tabular}{lcc}
        \toprule
        Category & Symbol & Description \\
        \midrule
        Spectroscopic diagnostics
        & $z_{\rm meas}$ & Best-fit redshift (fiducial) \\
        & $z_{\rm 2nd}$ & Second-best reliable redshift \\
        & ZWARNING & Redshift warning flag \\
        & $N_{\rm lines}$ & Number of matched emission lines \\
        & $\mathrm{C\,\textsc{iii}}]_{\rm snr}$ & SNR of the C\,\textsc{iii}] line fit \\
        & $[\mathrm{O\,\textsc{ii}}]_{\rm snr}$ & SNR of the [O\,\textsc{ii}] line fit \\
        & $\mathrm{H\gamma}_{\rm snr}$ & SNR of the H$\gamma$ line fit \\
        & $\mathrm{H\beta}_{\rm snr}$ & SNR of the H$\beta$ line fit \\
        & $[\mathrm{O\,\textsc{iii}}]_{\rm snr}$ & SNR of the [O\,\textsc{iii}] line fit \\
        & $\mathrm{H\alpha}_{\rm snr}$ & SNR of the H$\alpha$ line fit \\
        \midrule
        Photometric properties
        & $NUV$ & $NUV$-band magnitude \\
        & $u$ & $u$-band magnitude \\
        & $g$ & $g$-band magnitude \\
        & $r$ & $r$-band magnitude \\
        & $i$ & $i$-band magnitude \\
        & $z$ & $z$-band magnitude \\
        & $y$ & $y$-band magnitude \\
        & $GU$ & $GU$-band magnitude (slitless) \\
        & $GV$ & $GV$-band magnitude (slitless) \\
        & $GI$ & $GI$-band magnitude (slitless) \\
        & $n$ & Sérsic index \\
        & $R_e$ & Effective radius (pix) \\
        & $b/a$ & Axis ratio \\
        & PA & Position angle (deg) \\
        \bottomrule
    \end{tabular}
    \begin{minipage}{\linewidth}
    \textbf{Note.} All spectroscopic diagnostics extracted from the redshift measurement process correspond to the fiducial redshift ($z_{\mathrm{meas}}$), with the exception of $z_{\mathrm{2nd}}$. Photometric properties are primarily from CSST imaging, with the $GU$, $GV$, and $GI$ bands obtained from slitless spectra.
    \end{minipage}
\end{table*}

\section{Methodology}
\label{sec:method}
Our objective is to select a high-purity subset of galaxies from the CSST slitless spectroscopic survey while maximizing completeness.
This task presents two interconnected challenges: processing the high-dimensional, nonlinear relationships within the photometric and spectroscopic features, and scaling efficiently to the $\sim$62 million galaxies in our parent sample.
To address these, we employ the eXtreme Gradient Boosting (XGBoost) algorithm \citep{Chen:2016aa}.
XGBoost is a state-of-the-art implementation of gradient-boosted decision trees, renowned for its predictive accuracy, computational efficiency, and robustness against overfitting \citep{Shwartz-Ziv:2022aa,Grinsztajn:2022aa}.
Critically, it natively handles missing data by internally optimizing decision splits to maximize information gain, which is essential for processing real-world observational datasets that commonly contain incomplete entries due to measurement noise or instrument limitations.
Additionally, its tree-based architecture can provide inherent interpretability through feature importance metrics and decision pathways, and it can be interpreted seamlessly with SHAP \citep{Lundberg:2017aa} to deliver both global and local interpretability, potentially offering physical insights into the model's predictions.
This combination of capabilities makes it exceptionally suitable for our large-scale classification task.

\subsection{Model Training}

Prior to model training, the feature matrix was preprocessed to ensure numerical stability.
Non‑numerical entries (NaN or infinite values) in the feature matrix were replaced with a constant sentinel value of $-999$, and all features were subsequently standardized to zero mean and unit variance using a \texttt{StandardScaler}.
For binary classification, each galaxy was labeled as accurate (positive class, yields 1) if its redshift measurement satisfied $|\Delta z| / (1+z_{\rm true}) \leqslant 0.002$, and inaccurate (yields 0) otherwise.

We employed an XGBoost classifier configured with the logistic loss objective (\texttt{binary:logistic}).
To optimize model performance, a randomized hyperparameter search was conducted over 50 iterations using 3-fold cross-validation on the training set.
The search space for key hyperparameters included: the number of estimators (\texttt{n\_estimators}: 300--1000), learning rate (\texttt{learning\_rate}: 0.01--0.11), maximum tree depth (\texttt{max\_depth}: 4--10), and subsample ratios (\texttt{subsample} and \texttt{colsample\_bytree}, both ranging from 0.7 to 1.0).
The L1 and L2 regularization strengths, \texttt{reg\_alpha} and \texttt{reg\_lambda}, were sampled from the intervals [0, 1] and [0, 2], respectively.
The area under the receiver operating characteristic (ROC) curve (AUC) was used as the primary optimization metric.

The final model was trained using the optimal hyperparameter configuration identified by the search; for the key parameters tuned, the values were \texttt{n\_estimators} = 886, \texttt{learning\_rate} = 0.0392, \texttt{max\_depth} = 9, \texttt{subsample} = 0.958, \texttt{colsample\_bytree} = 0.742, \texttt{reg\_alpha} = 0.983, and \texttt{reg\_lambda} = 0.934.
Early stopping was applied, monitored on the separate validation set, to halt training if the validation log‑loss showed no improvement for 50 consecutive rounds.
This approach effectively determined the optimal number of boosting iterations (865) within the predefined \texttt{n\_estimators} range.
Model discrimination performance was evaluated via the ROC curve and its AUC on the validation set, which also informed the subsequent selection of the classification threshold.

All analyses were performed in Python 3.9.23 using NumPy \citep[1.26.4;][]{Harris:2020aa}, scikit‑learn \citep[1.6.1;][]{JMLR:v12:pedregosa11a}, XGBoost \citep[2.1.4;][]{Chen:2016aa}, and SHAP \citep[0.45.1;][]{Lundberg:2017aa}.
Model training leveraged parallel processing across all available CPU cores to accelerate the hyperparameter search and boosting iterations.
Owing to the computational efficiency of XGBoost, the entire training process completed within minutes on a node equipped with 72 CPU cores.
The complete workflow, from data preprocessing to model evaluation, was implemented in a modular pipeline that automatically caches intermediate results (e.g., standardized features, data splits) to facilitate reproducibility and efficient experimentation.

\subsection{Threshold Selection Strategy}\label{threshold_selection}
\begin{figure}
    \centering
	\includegraphics[width=8cm]{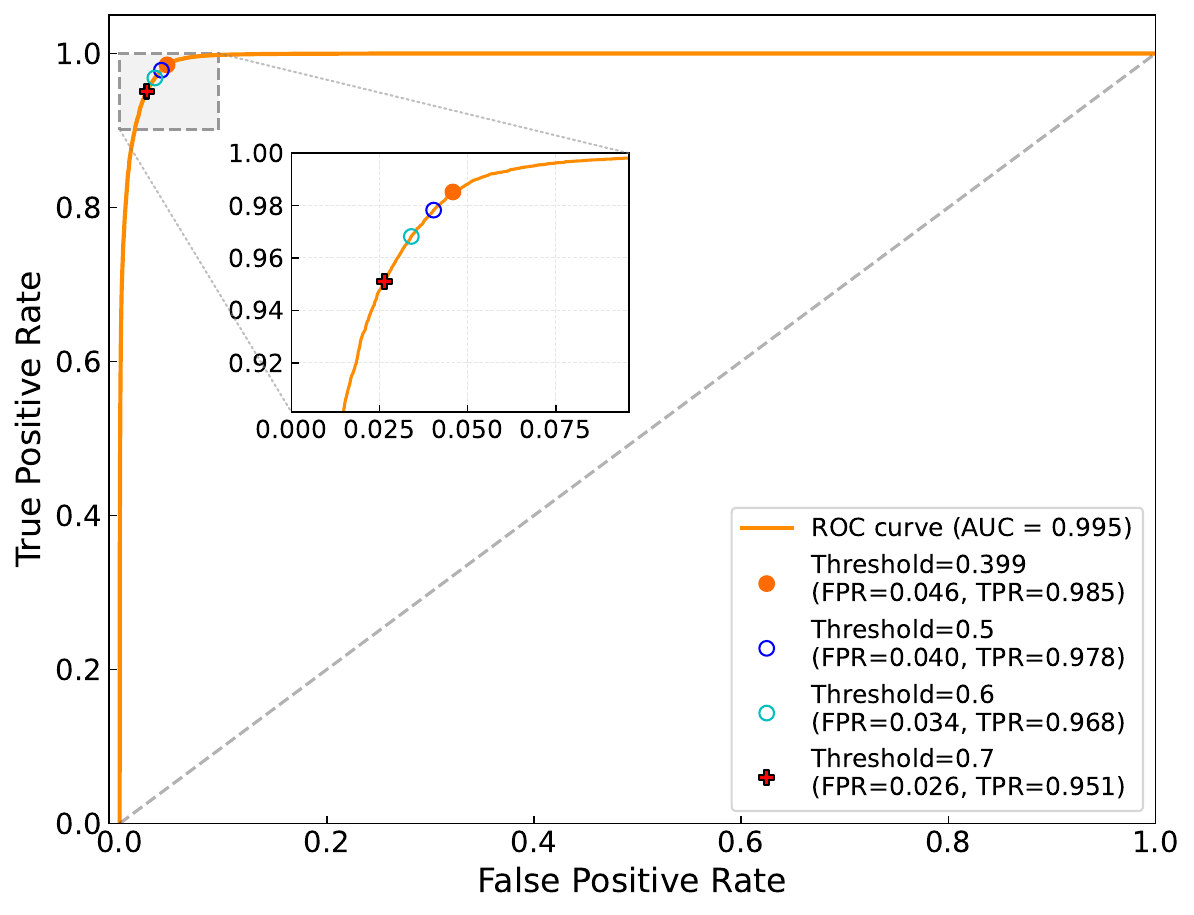}
    \caption{Receiver operating characteristic (ROC) curve of the model on the validation set. The area under the curve (AUC) is shown. The dashed diagonal line indicates random classification performance (TPR = FPR). The optimal threshold based on Youden's index (maximizing $\text{TPR}-\text{FPR}$) is shown as a solid orange circle. The blue and cyan hollow circles denote the default and a referenced threshold, respectively. The final selected threshold is indicated by a filled plus marker. The inset shows a magnified view of the region containing these threshold markers.}
    \label{fig:roc_curve_validation}
\end{figure}

Upon completion of model training, we selected an appropriate probability threshold on the validation set to ensure the purity of the final galaxy sample.
This post hoc tuning balances the trade-off between sample completeness and contamination control, as visualized by the ROC curve in Figure~\ref{fig:roc_curve_validation}.
In this framework, the true positive rate (TPR; equivalent to recall) reflects the fraction of accurately measured galaxies correctly retained in the catalog, while the false positive rate (FPR) corresponds to the fraction of misidentified galaxies erroneously included.
The diagonal dashed line indicates the performance of a random classifier (AUC = 0.5), whereas an ideal classifier would achieve AUC = 1.

Our model attains an AUC of 0.995, demonstrating excellent separability.
We evaluated four representative thresholds (annotated in Figure~\ref{fig:roc_curve_validation}):
\begin{itemize}
    \item Youden's index optimum (0.399): Maximizes $\mathrm{TPR} - \mathrm{FPR}$.
    \item Default threshold (0.5).
    \item Reference threshold (0.6).
    \item Selected threshold (0.7).
\end{itemize}
Their corresponding performance metrics are: Youden's optimum (TPR = 0.985, FPR = 0.046), default 0.5 (TPR = 0.978, FPR = 0.040), reference 0.6 (TPR = 0.968, FPR = 0.034), and our selected 0.7 (TPR = 0.951, FPR = 0.026).

Although Youden's optimum provides the best overall balance, we chose the higher threshold of 0.7 for two primary reasons.
First, it reduces the FPR by 43\% (from 0.046 to 0.026), significantly lowering potential interloper contamination while maintaining a high TPR of 0.951.
Second, constructing a high-purity catalog is paramount for cosmological studies with CSST slitless spectroscopy, as even moderate contamination can systematically bias measurements of large-scale structure.
This selection prioritizes purity, aligning with the specific scientific requirements of the analysis.

\section{Results}\label{sec:results}

We present the performance of the high-purity redshift catalog produced by our XGBoost framework.
The finalized model, which was trained on the training set with optimal hyperparameters and uses the classification threshold of 0.7 selected on the validation set, was first applied to the held-out test set, and subsequently to the parent sample, to provide an unbiased evaluation and demonstrate scalability.
Unless otherwise noted, all performance metrics and SHAP-based interpretability analyses are reported on the test set.

Section~\ref{sec:model_performance} summarizes the key catalog metrics on the test and parent samples: selection efficiency, fraction of accurate measurements, and outlier fraction.
Section~\ref{sec:feature_importance} presents the model's decision mechanism via SHAP analysis, and Section~\ref{sec:comparative_analysis} evaluates the performance under simplified feature configurations.

\subsection{Performance on Test and Parent Sample}\label{sec:model_performance}

\begin{figure*}
\centering
    \begin{minipage}{8cm}
        \centering
        \includegraphics[height=8.5cm]{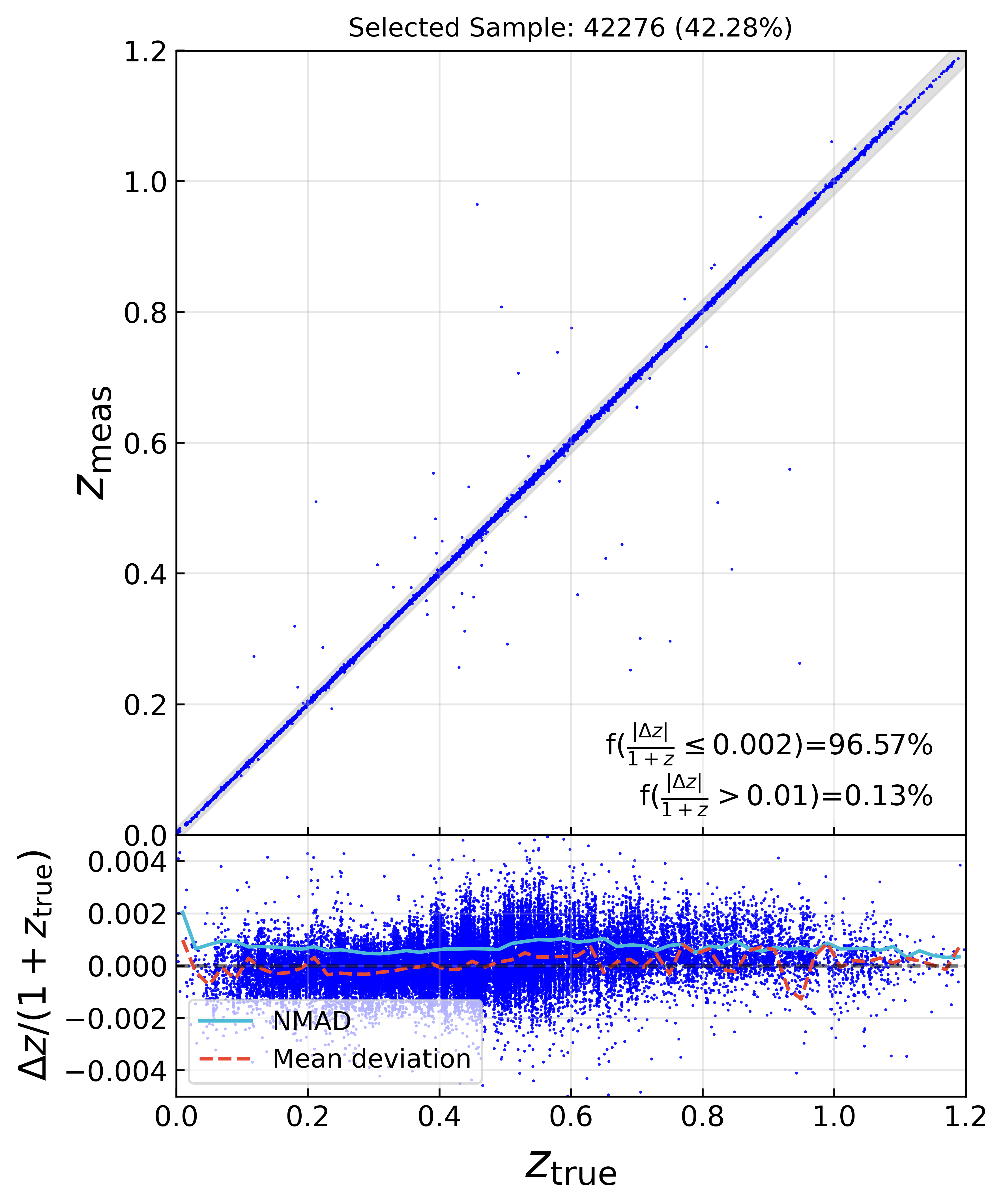}
    \end{minipage}
	\begin{minipage}{8cm}
    	\centering
        \includegraphics[height=8.5cm]{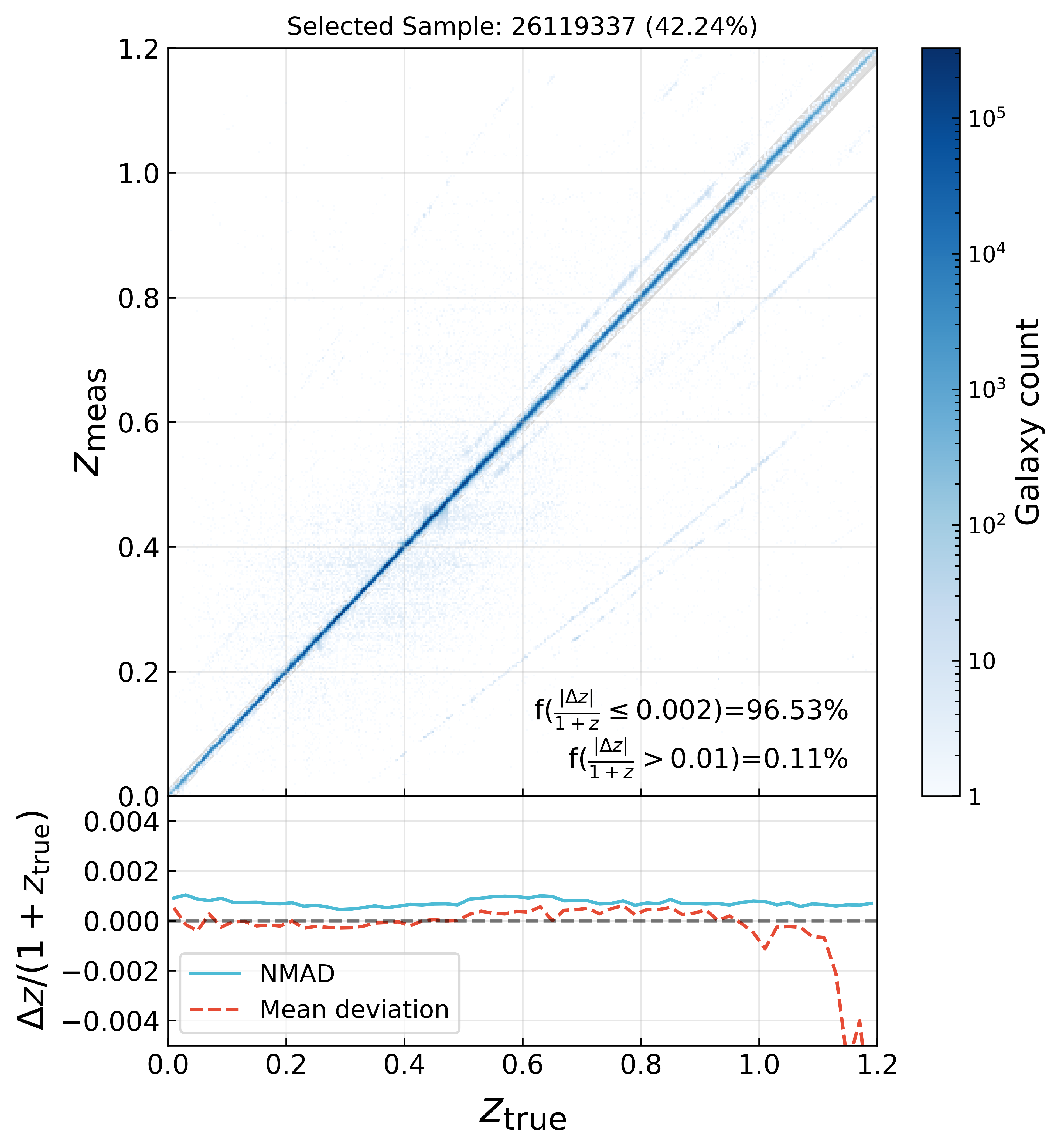}
    \end{minipage}
    \caption{Comparison of model performance on the test sample (left) and the parent sample (right). The upper subplot in each panel shows the measured redshift plotted against the true redshift for the corresponding sample. The lower subplots present the redshift offset, $\Delta z /(1+z_{\mathrm{true}})$, as a function of $z_{\mathrm{true}}$; this detailed offset analysis is only available for the test sample (left), due to the prohibitive size of the parent dataset. In both lower subplots, the mint‑blue solid line and coral dashed line represent the $\sigma_{\mathrm{NMAD}}$ scatter and the binned mean deviation, respectively. The slight increase in the mean deviation at $z > 1.1$ in the parent sample is caused by the upper limit of the measured redshift ($z_{\mathrm{meas}} \lesssim 1.2$), which truncates the error distribution and biases the mean toward negative values. This effect is absent in the smaller test set due to the low occurrence rate of such boundary outliers.
    }
    \label{fig:model_performance}
\end{figure*}

We evaluate the model's classification performance using three key metrics:
\begin{itemize}
\item The selection efficiency, defined as the fraction of selected objects relative to the total sample.
\item The fraction of accurate measurements ($f_{\rm accurate}$), defined as the fraction of selected objects satisfying $|\Delta z| \leqslant 0.002(1+z_{\rm{true}})$.
\item The outlier fraction ($f_{\rm outlier}$), defined as the fraction of selected objects satisfying $|\Delta z| > 0.01(1+z_{\rm{true}})$.
\end{itemize}

Additionally, to comprehensively assess the redshift quality, we compute the normalized median absolute deviation ($\sigma_{\rm NMAD}$), defined as 1.4826 $\times$ Median$(| \frac{\Delta z}{1+z_{\rm true}} - {\rm Median}(\frac{\Delta z}{1+z_{\rm true}}) |)$, the mean deviation (bias), and the fraction of catastrophic interlopers ($f_{\rm cat}$), defined as objects with $|\Delta z| > 0.1(1+z_{\rm{true}})$.

Figure~\ref{fig:model_performance} compares the performance on the test set and the parent sample.
The classifier shows nearly identical selection efficiencies of 42.3\% (test) and 42.2\% (parent sample), demonstrating remarkable stability.
This selection efficiently recovers the population of galaxies with inherently accurate redshift measurements, which constitutes 42.9\% of the parent sample (Figure~\ref{fig:sample_cut_compare}).
The upper panels confirm the excellent quality of the retained redshifts, showing tight agreement between the measured and true values.
Crucially, the purification is highly effective: 96.6\% (test) and 96.5\% (parent sample) of the selected galaxies have accurate measurements, while the outlier fraction is suppressed to merely 0.13\% and 0.11\%, respectively.
The final catalog achieves a low redshift uncertainty ($\sigma_{\mathrm{NMAD}} = 0.0007$) and a minimal catastrophic interloper fraction ($f_{\mathrm{cat}} \simeq 0.05\%$).
For comparison, the fractions of catastrophic interlopers for the `SNR Cut' and `Additional Cuts' cases shown in Figure~\ref{fig:sample_cut_compare} are 4.81\% and 2.49\%, respectively.
The lower panels provide a detailed, binned view of the residual $\Delta z / (1+z_{\mathrm{true}})$, revealing its small, redshift-independent mean bias and scatter ($\sigma_{\mathrm{NMAD}}$), which corroborate the outstanding global metrics.

We note that the residual off-diagonal linear aggregations visible in the right panel of Figure~\ref{fig:model_performance} originate from a small population of emission-line confusion interlopers that survive the fiducial probability threshold. Although applying additional spectroscopic quality cuts can further suppress these residuals, doing so substantially reduces the overall sample size without a commensurate gain in purity, given that the outlier fraction is already at the $0.1\%$ level. The framework therefore prioritizes the retention of valuable galaxies while maintaining purity well within the requirements of large-scale structure studies.

\subsection{Feature Importance Analysis with SHAP}\label{sec:feature_importance}
\begin{figure}[htbp]
\centering
    \includegraphics[width=8cm]{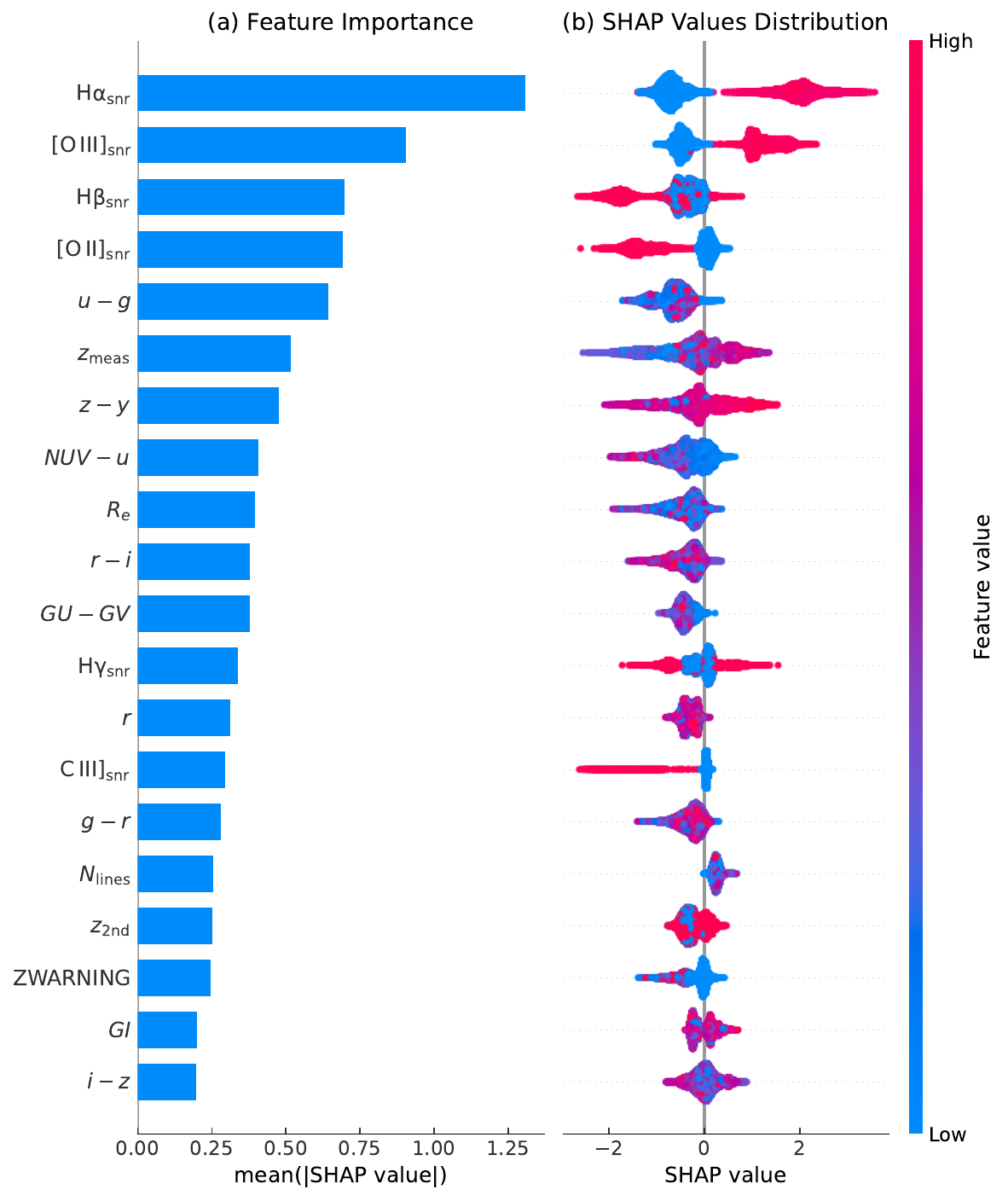}
    \caption{SHAP visualizations for the top 20 most influential features: (a) Feature importance ranking based on mean absolute SHAP values, quantifying average magnitude of feature contributions. (b) Beeswarm plot displaying the distribution of SHAP values for each feature, where dot color indicates feature value (red=high and blue=low), horizontal position shows impact direction (right=positive and left=negative), and density reveals value distribution patterns.
    }
    \label{fig:feature_importance}
\end{figure}

To interpret our classifier and quantify feature importance, we employ SHAP analysis.
As a unified framework rooted in cooperative game theory, SHAP quantifies each feature's contribution by computing its average marginal effect across all possible feature subsets.
This approach offers a more complete and theoretically grounded understanding than heuristic importance metrics native to XGBoost (e.g., gain, cover, and frequency), which often fail to account for feature interactions.
SHAP values are both consistent and additive, ensuring a faithful attribution of influence.
As a result, SHAP enables us not only to rank features by overall impact but also to examine the direction and magnitude of each feature's effect on specific predictions, thereby revealing the physical and diagnostic cues the model uses to distinguish accurate redshifts from interlopers.

Figure~\ref{fig:feature_importance} presents the SHAP analysis for the top 20 most influential features, which account for 87.7\% of the cumulative absolute SHAP value.
The bar plot (a) ranks individual features by their mean absolute SHAP value.
From this ranking, we find that both spectroscopic diagnostics (particularly emission-line SNRs) and photometric properties (particularly colors) are the influential features for the model's decisions.
Besides, the ranking shows that the photometric properties from the three broad slitless bands ($GU$, $GV$, and $GI$) can be safely omitted when they are lacking because the seven narrow bands collectively cover the same wavelength range and provide sufficient information.
The beeswarm plot (b) further elucidates how these features drive predictions: each point represents a galaxy, colored by its feature value (red: high and blue: low) and positioned by its SHAP value.
This visualization demonstrates the model's capacity to capture high-dimensional, nonlinear relationships within the feature space, effectively synthesizing diverse information for robust classification.

A closer inspection of the beeswarm plot provides further nuance regarding feature reliability.
First, the SNRs of $\mathrm{H\alpha}$ and $[\mathrm{O\,\scriptstyle{III}}]$ show a consistent, reliable association with accurate classification.
In contrast, some ostensibly strong spectral signals, such as those attributed to the $[\mathrm{O\,\scriptstyle{II}}]$ and $\mathrm{C\,\scriptstyle{III}}]$ emission lines, likely correspond to spurious features rather than genuine physical properties, since they are not prominent enough in the observed spectra \citep[see Figure 3 in][]{Sui:2025aa}.
For $\mathrm{H\beta}$ and $\mathrm{H\gamma}$, however, the picture is more complex: high SNR acts as an ambivalent predictor, associated with both correct and incorrect identifications. 
Notably, the distributions of colors and magnitudes exhibit clear non-linear relationships with the SHAP values, a pattern that is physically well-motivated and analogous to the complex correlations leveraged in photometric redshift estimation.
This reinforces the interpretation that the classifier is learning physically grounded relationships from the data.

\subsection{Performance of Simplified Feature Configurations}\label{sec:comparative_analysis}

\begin{figure*}
\centering
    \begin{minipage}{8cm}
        \centering
        \includegraphics[width=8cm]{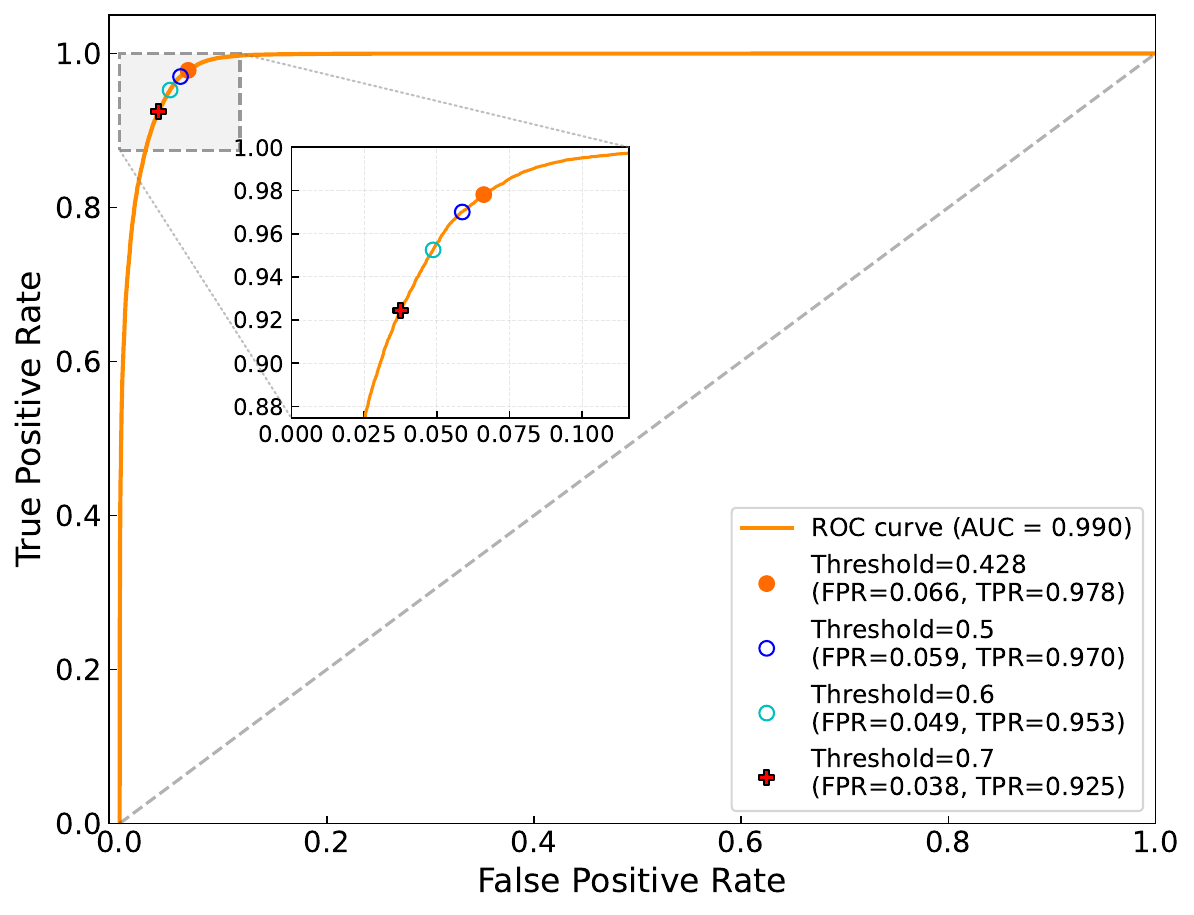}
    \end{minipage}
    \hspace{0.2cm}
	\begin{minipage}{8cm}
    	\centering
        \includegraphics[width=8cm]{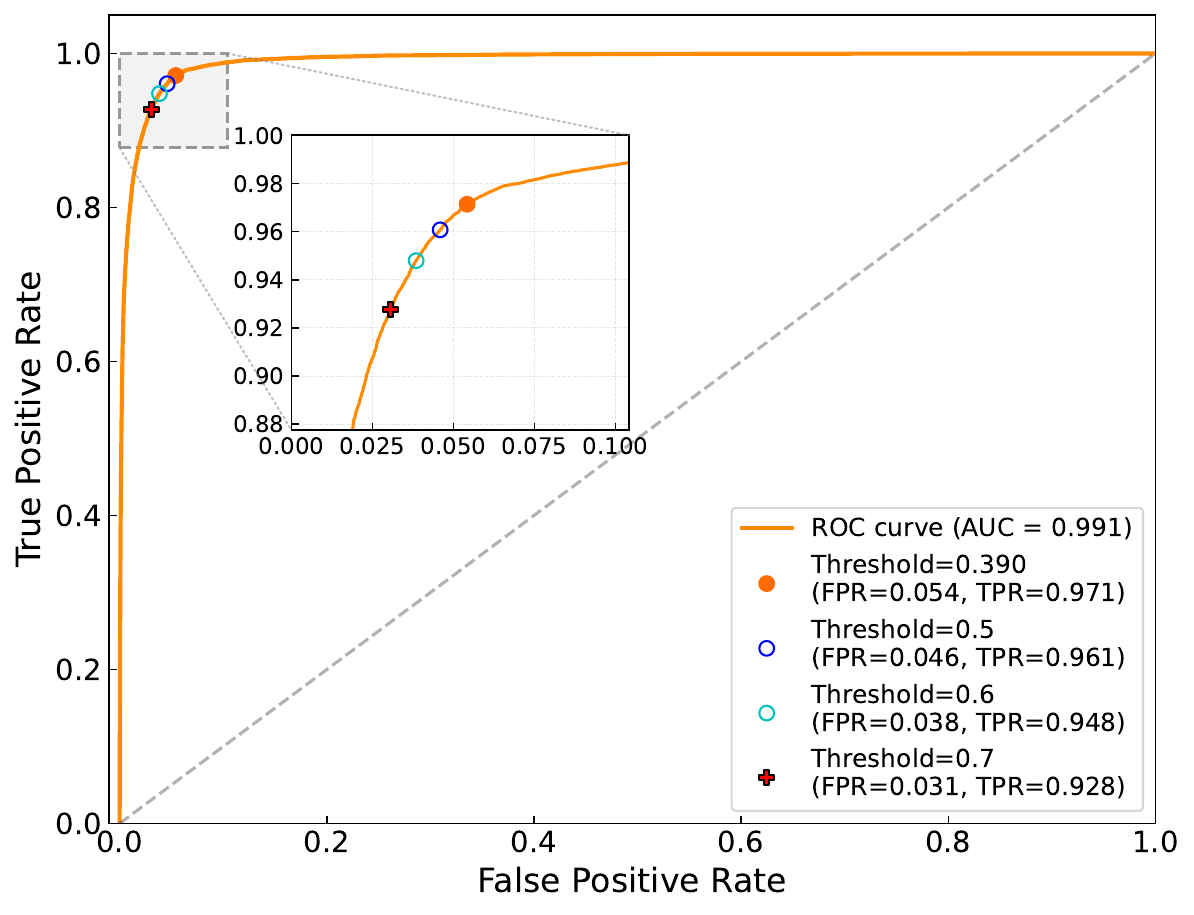}
    \end{minipage}
    \caption{ROC curves of the simplified models evaluated on the test set. Left: model using only photometric data and measured redshift $z_{\rm meas}$. Right: model using only spectroscopic diagnostics.}
    \label{fig:roc_curve_simplified_feature}
\end{figure*}

\begin{figure*}
\centering
    \begin{minipage}{8cm}
        \centering
        \includegraphics[width=8cm]{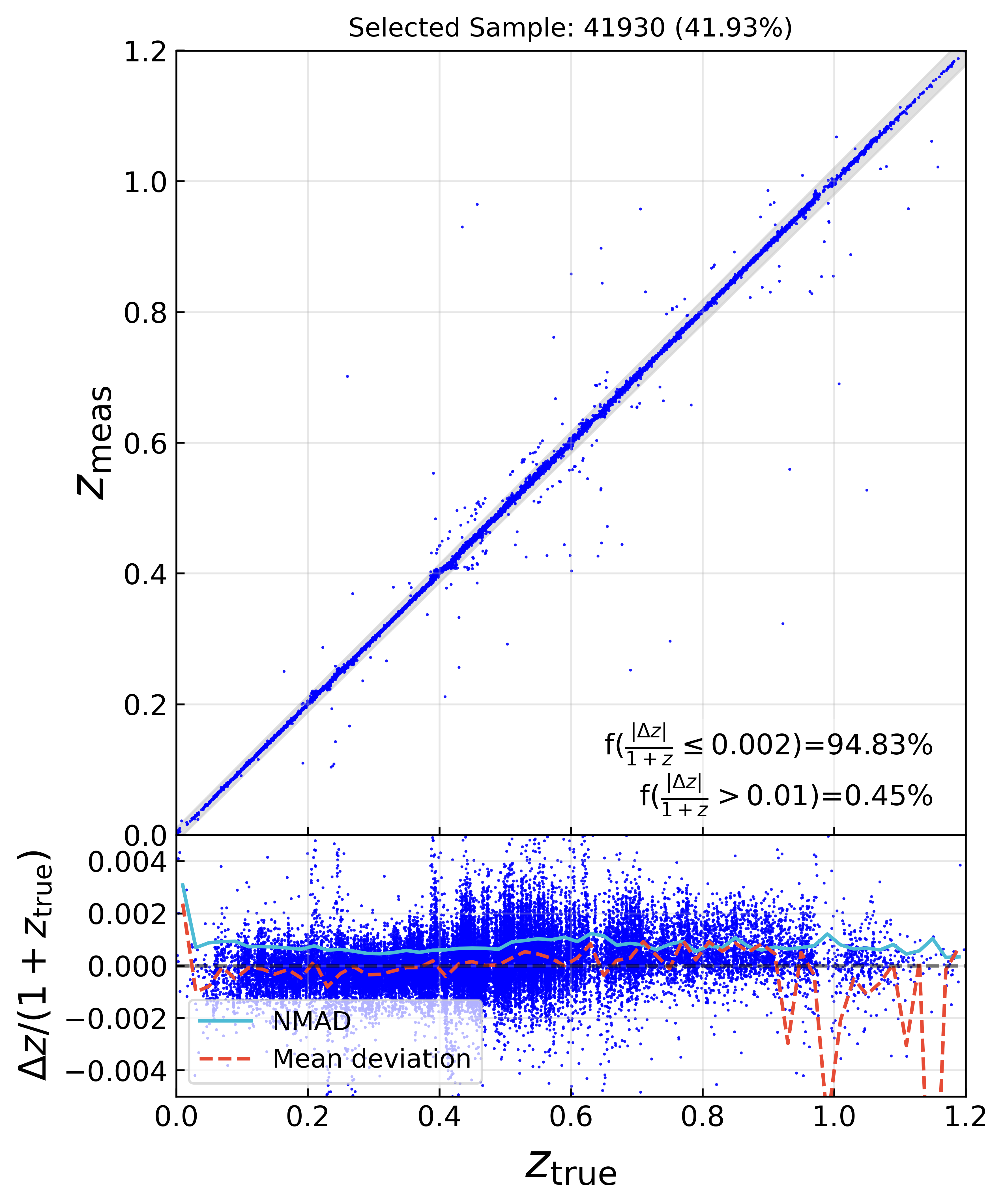}
    \end{minipage}
    \hspace{0.2cm}
	\begin{minipage}{8cm}
    	\centering
        \includegraphics[width=8cm]{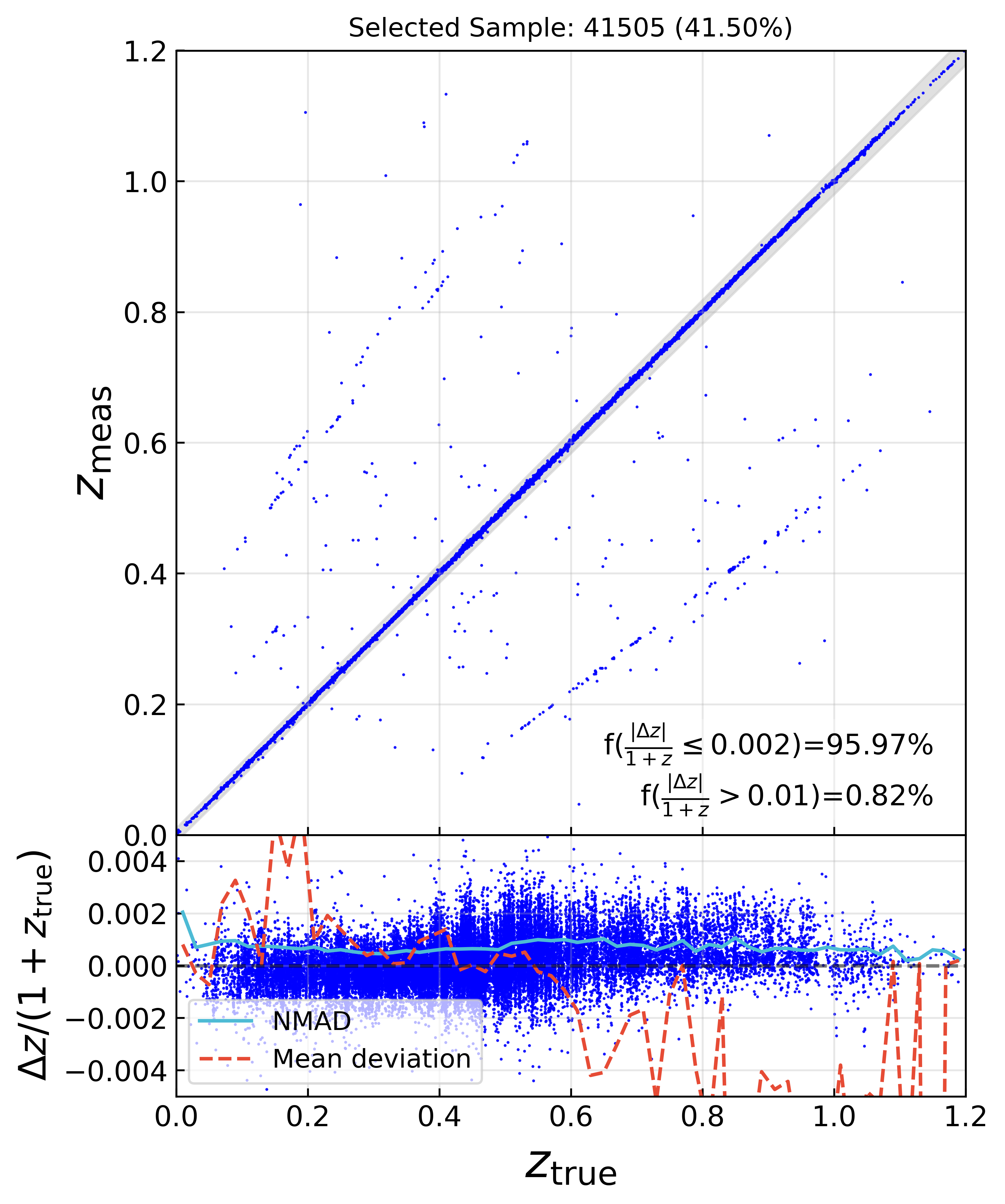}
    \end{minipage}
    \caption{Performance of the simplified models at the fiducial threshold of 0.7. The panel layout and feature configurations follow those described in Figure~\ref{fig:roc_curve_simplified_feature}.}
    \label{fig:simplified_model_performance}
\end{figure*}

\begin{table*}[htbp]
\centering
\caption{Performance comparison on the test set using the fixed threshold of 0.7.}
\label{tab:performance_comparison}
\begin{tabular}{lccccc}
\toprule
Model Configuration & AUC & Selection Efficiency & $f_{\rm accurate}$ & $f_{\rm outlier}$ & $f_{\rm cat}$ \\
& &(\%)&(\%)&(\%)&(\%)\\
\midrule
Fiducial (all features) & 0.995 & 42.28 & 96.57 & 0.13 & 0.05\\
Photometry + $z_{\rm meas}$ & 0.990 & 41.93 & 94.83 & 0.45 & 0.07\\
Diagnostics only & 0.991 & 41.50 & 95.97 & 0.82 & 0.68\\
\bottomrule
\end{tabular}
\begin{minipage}{\linewidth}
    \textbf{Note.} The fiducial model combines all available features, while the simplified models use subsets.
    \end{minipage}
\end{table*}

Beyond the fiducial model incorporating all available features, we conduct an ablation study to assess the contribution of different data modalities.
Two simplified configurations are examined: (i) using photometric data and the measured redshift (23 features: photometry + $z_{\rm meas}$), and (ii) using only spectroscopic diagnostics (10 features).
Both simplified models are trained, validated, and tested on the same data splits as the fiducial model to ensure a controlled comparison.
Crucially, all configurations are evaluated using the same probability threshold of 0.7, which was optimized for the fiducial model on the validation set.
No additional threshold tuning was performed, ensuring that performance differences arise solely from the feature sets.

The ROC curves of the simplified models are presented in Figure~\ref{fig:roc_curve_simplified_feature}, and their concrete performance at the fiducial threshold of 0.7 is shown in Figure~\ref{fig:simplified_model_performance}. Corresponding quantitative metrics are summarized in Table~\ref{tab:performance_comparison}.
Despite achieving comparable selection efficiency and accurate measurement fractions (within $\sim$1–2 percentage points of the fiducial model), both simplified configurations exhibit substantially higher contamination.
The outlier fraction ($f_{\rm outlier}$) increases 3.5-fold (to 0.45\%) for the photometry-only model and 6.3-fold (to 0.82\%) for the diagnostics-only model.
More critically, in the diagnostics-only configuration, the catastrophic interloper fraction ($f_{\rm cat}$) shows a marked 13.6-fold increase from 0.05\% to 0.68\%.
Such catastrophic misidentifications would introduce significant systematic biases in downstream analyses, as evidenced by the increased mean redshift deviation (lower-right panel of Figure~\ref{fig:simplified_model_performance}).

The pronounced degradation in purity when using only spectroscopic diagnostics, contrasted with the relatively stable performance of the photometry-only model, underscores the unique value of multiband photometry in identifying and filtering interlopers within slitless spectroscopic data.
While spectroscopic diagnostics assess internal consistency, photometry provides an independent prior on galaxy spectral energy distributions, offering complementary constraints that are essential for constructing high-fidelity redshift catalogs.

\section{Conclusions}\label{sec:conclusions}
This work presents an XGBoost-based classifier designed to filter interlopers from slitless spectroscopic catalogs by synergistically combining multiband photometry with spectroscopic diagnostics.
Using simulated CSST slitless spectroscopy data, we demonstrate that the model robustly achieves an optimal balance between high purity and high completeness.
Operating at a fixed probability threshold of 0.7, the model delivers strong performance: it attains a $\sim$96.6\% fraction of accurate measurements and an outlier fraction of only 0.13\% within the selected sample, while recalling over 95\% of galaxies with accurate redshifts overall.
Interpretability analysis via SHAP confirms the physical consistency of the model’s decisions, clearly identifying the most influential photometric and diagnostic features.

Critically, an ablation study underscores the indispensable role of photometric data.
Removing photometric features leads to a 6.3-fold increase in the overall outlier fraction and a 13.6-fold surge in catastrophic interlopers, highlighting that photometry provides an independent constraint crucial for identifying severe misidentifications.
Conversely, a model relying solely on photometry combined with measured redshift maintains relatively stable performance, reinforcing its unique utility in interloper screening.

The implications of this method extend beyond sample purification.
The predictive framework and identified salient features could inform the redshift measurement process itself, particularly for challenging cases such as galaxies with only a single detected emission line \citep{Leung:2017aa,Davis:2023aa,Mentuch-Cooper:2023aa}, a scenario that lies beyond the capabilities of the current CSST measurement algorithm (as it requires at least two matched lines).
Here, photometric attributes and spectral characteristics such as the line profile shape (e.g., full width at half-maximum) and relative strength (e.g., equivalent width) may serve as vital priors to break degeneracies, potentially increasing the yield of reliable redshifts.

We acknowledge several avenues for future work.
First, while our analysis is based on simulated data with a defined feature set, applying the framework to mock data with more complex observational effects \citep{Wei:2025aa,Zhang:2025ae} and, ultimately, real observational data (e.g., from Euclid or future CSST releases) is essential.
This will validate performance under realistic conditions and may unlock a richer set of exploitable spectroscopic diagnostics not fully captured in this analysis.
Second, constructing an optimally balanced training set from real surveys presents a practical challenge that could introduce selection effects.
Given the relatively low redshift coverage ($0<z<1$) of CSST slitless spectroscopy, this challenge could be addressed by leveraging data from deep surveys and multiple ground-based surveys to increase sample diversity and representation, thereby training more robust and accurate models.
Third, exploring the model’s adaptability to other slitless spectroscopic surveys (e.g., Euclid and Roman) would demonstrate its generalizability.

In conclusion, this study establishes a machine learning framework that effectively leverages the complementary strengths of photometry and spectroscopy.
By simultaneously ensuring high purity and completeness, this framework provides a robust pathway to generating the high-fidelity, large-scale redshift catalogs essential for unlocking the full cosmological potential of CSST and similar slitless spectroscopic surveys.

Key code and dataset presented in this work are publicly available on Zenodo \citep[doi: \href{https://doi.org/10.5281/zenodo.19632048}{10.5281/zenodo.19632048};][]{peng_2026_dataset}.
The complete dataset, incorporating data from \cite{Wen:2024ab} and \cite{Sui:2025aa}, is available from the authors upon reasonable request.

\begin{acknowledgments}
This work was supported by the National Key R\&D Program of China (Nos.\ 2023YFA1607800 and 2023YFA1607802), the National Natural Science Foundation of China (grant Nos.\ 12595310 and 12273020), the China Manned Space Project with No.\ CMS-CSST-2025-A04, the “111” Project of the Ministry of Education under grant No.\ B20019, and the sponsorship from Yangyang Development Fund.
This project is supported in part by Office of Science and Technology, Shanghai Municipal Government (grant Nos.\ 24DX1400100 and ZJ2023-ZD-001).
This work made use of the Gravity Supercomputer at the Department of Astronomy, Shanghai Jiao Tong University.
\end{acknowledgments}


\bibliography{sample701}{}
\bibliographystyle{aasjournalv7}



\end{document}